\documentstyle[11pt,newpasp,twoside,epsf]{article}
\newcommand{\civ}{\mbox{C{\footnotesize\hspace{0.5mm}IV}}}
\newcommand{\siiv}{\mbox{Si{\footnotesize\hspace{0.5mm}IV}}}
\newcommand{\nv}{\mbox{N{\footnotesize\hspace{0.5mm}V}}}
\newcommand{\oiv}{\mbox{O{\footnotesize\hspace{0.5mm}IV}}}
\newcommand{\ov}{\mbox{O{\footnotesize\hspace{0.5mm}V}}}
\newcommand{\hei}{\mbox{He{\footnotesize\hspace{0.5mm}I}}}

\def\edcomment#1{\iffalse\marginpar{\raggedright\sl#1\/}\else\relax\fi}
\reversemarginpar

\begin{document}
\title{{STIS} Observations of five hot DA white dwarfs}
 \author{N P Bannister, M A Barstow}
\affil{Department of Physics and Astronomy, University of
Leicester, University Road, Leicester, LE1 7RH, U.K.}
\author{J B Holberg}
\affil{Lunar and Planetary Laboratory, University of Arizona,
Tucson, AZ 85721, USA. }
\author{F C Bruhweiler}
\affil{Institute for Astrophysics and Computational Sciences,
Department of Physics, The Catholic University Of America,
Washington, D.C. 20064, USA.}

\begin{abstract}
We present some early results from a study of five hot DA white
dwarf stars, based on spectra obtained using STIS. All show
multiple components in one or more of the strong resonance
absorption lines typically associated with the stellar photosphere
(e.g. \civ, \siiv, \nv\ and \ov). Possible relationships between
the non-photospheric velocity components and the interstellar
medium or local stellar environment, are investigated, including
contributions from gravitational redshifting.
\end{abstract}

\section{Introduction}
Several studies of hot DA white dwarf stars have revealed highly
ionized species at non-photospheric velocities. While shifted
features in the spectrum of Feige 24 could be explained by the
binary nature of this system (Dupree \& Raymond, 1982), {\em IUE}
observations of the isolated DA white dwarf CD -38\deg10980
revealed Si and C absorption features at velocities displaced by
-12 km/s with respect to the photospheric velocity. These features
have been explained in terms of a dense gaseous halo in close
proximity to the star (Holberg, Bruhweiler \& Andersen (1995)).

We have examined high resolution EUV spectra of five hot (T$_{eff}
> 54,000$
 K) DA white dwarfs, 
listed with their basic physical parameters in table 1. All show
multiple components in the \civ\ resonance doublet, and some
exhibit features in other lines. We compare the velocity of each
component to that of the line-of-sight ISM lines observed in the
spectra, and to the velocity of the local interstellar cloud
(LIC). Using stellar models, gravitational redshifts are estimated
to determine whether these shifted components could be produced by
material within the gravitational well of the star, and column
density estimates are derived to provide further information on
the nature of the material. In the case of WD 2218+706, a possible
link to a large molecular complex is suggested.

\begin{table}
\caption{Physical parameters of the five stars
included in this study}
\begin{tabular}{l c c c r r}
\tableline
 Star & T$_{eff}$ (K) & Mass (M\sun) & Radius (R\sun) &
Log g & Dist (pc) \\ \tableline G191-B2B & 54,000 & 0.51 & 0.020 &
7.5 & 69
\\ Feige 24 & 56,000 & 0.50 & 0.018 &
$>7.5$ & 70 \\ WD 2218+706 & 56,900 & 0.40 & 0.033 & 7.0 & 436 \\
REJ 0558+165 & 62,860 & 0.58 & 0.019 & 7.7 & 295 \\ REJ 1738+665 &
69,340 & 0.53 & 0.024 & 7.4 & 239 \\ \tableline
\end{tabular}
\end{table}

\section{Observations and data reduction}
All spectra were obtained using the STIS instrument on HST,
configured with the E140-M grating, and cover the wavelength range
from 1150 \AA\ - 1750 \AA\ at a resolution of 0.042 \AA\ FWHM.
Spectra were reduced using the IACS pipeline. Line velocities were
measured using code written by
one of us (JBH). 
The velocities and equivalent widths of resonance lines showing
two or more components were estimated by performing multiple
gaussian fits to the data using the {\em DIPSO} package provided
by the UK Starlink project. The LIC velocity in the direction of
each star was estimated using the vector published by Lallement et
al. (1995), who estimated V$_{lic} \approx 26$ km/s towards $186\
l_{II}\  -16\ b_{II}$. Application of simple spherical
trigonometry provides the radial component of this motion in the
direction of each object. Gravitational redshift estimates were
determined using stellar models produced by Matt Wood (Wood,
1995), and curves of growth derived using code also written by
JBH.

\section{Results and discussion}
\subsection{RE 1738+665}
RE 1738+665 is the hottest DA white dwarf to be detected by {\em
ROSAT}, as described by Barstow et al. (1994). We determine a
photospheric velocity $V_{phot} \approx 30 \pm 1$ km/s based on
absorption features arising from Fe, Ni and O which show no
multiple components; interstellar lines indicate a line-of-sight
ISM velocity of $V_{ism} \approx -18 \pm 1$ km/s. The
line-of-sight velocity of the LIC is estimated to be $V_{LIC}
\approx -3.4$ km/s. Figure 1 (left) shows the \civ\ resonance
doublet in RE 1738, with shifted components at $-18.5\pm 0.5$ km/s
dominating the 30 km/s photospheric contribution.

Shifted features with similar velocities are determined for
several other species, although in each case the photospheric
component is dominant. The \siiv\ doublet shows clear evidence of
multiplicity, with non-photospheric velocity components observed
at -17.7 $\pm 0.7$ km/s. Viewed individually, the lines of the
\nv\ doublet (1238.821 and 1242.804 \AA) show no evidence of
companions, but co-addition of these features suggests an extra,
weak line at $-15.2 \pm 2$ km/s. The \oiv\ doublet (1338.612 and
1343.512 \AA) shows no additional features, but the \ov\ line at
1371.292 \AA\ is accompanied by a weak shifted component at $-18.7
\pm 0.5$ km/s.

\begin{figure}[t!]
\plottwo{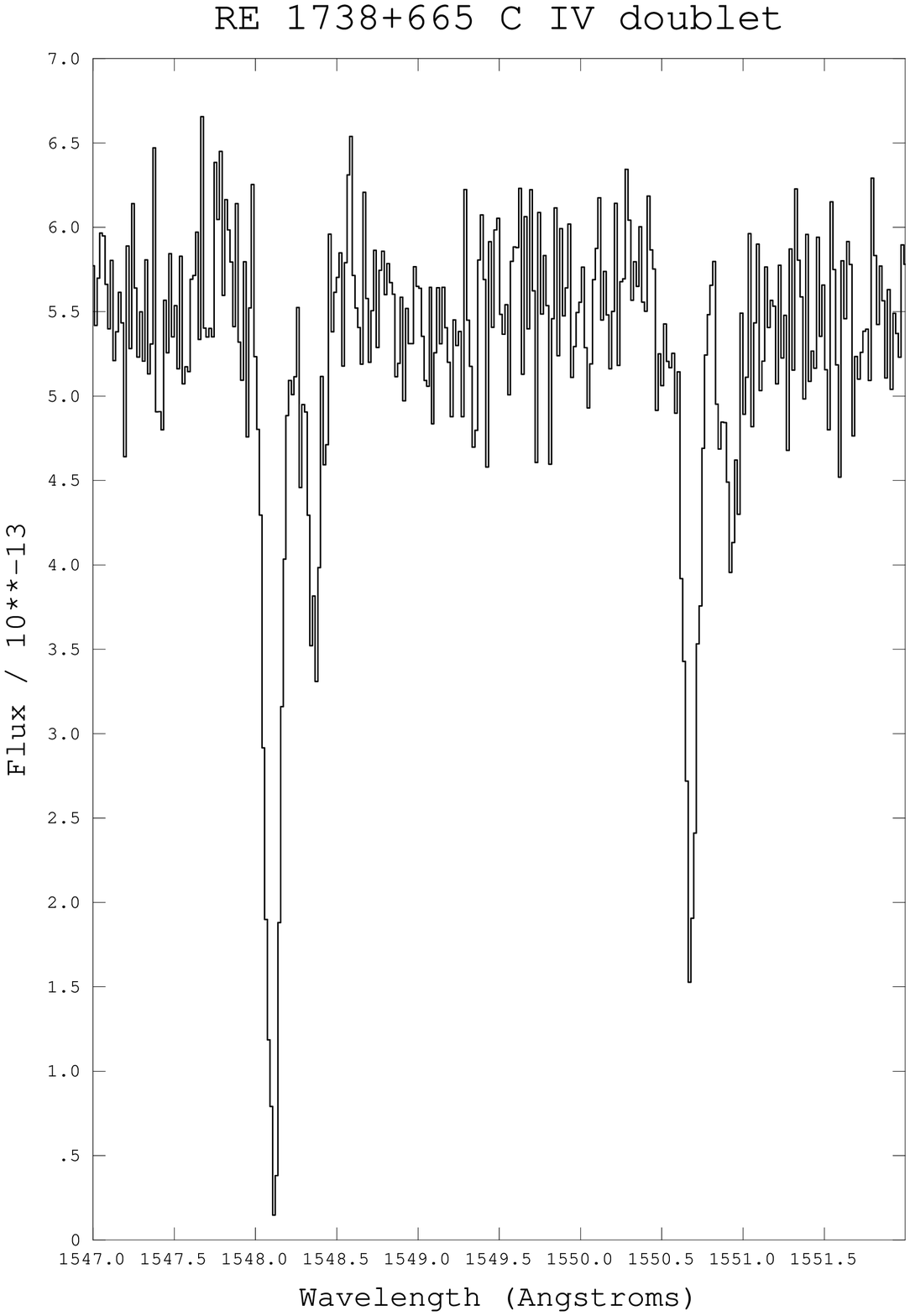}{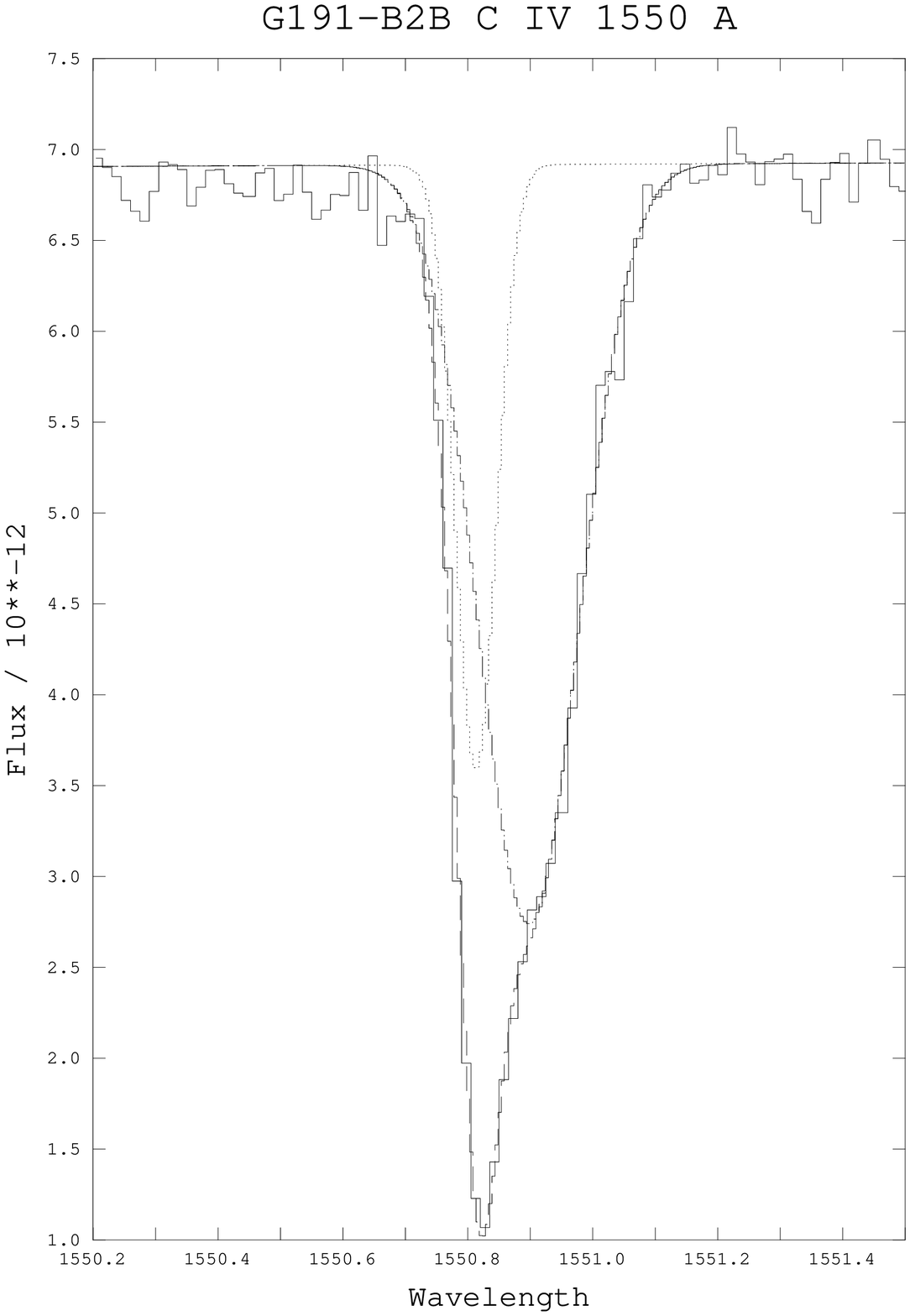} \caption{Velocity shifted,
high ionization components in two of the stars under discussion.
Left: clear multiplicity in the C IV doublet of RE 1738+665, with
the photospheric components dominated by the blueshifted lines.
Right: more subtle blending in the $\lambda$1550 line of G191-B2B,
shown with the compound Gaussian fit and its two components.}
\end{figure}

Our estimation of the gravitational redshift of this star is
$V_{gr} \approx 14$ km/s. Given uncertainties in this estimate and
in the velocity of the shifted components, $V_{gr}$ is comparable
to the velocity of these components, suggesting that the material
does not reside within the gravitational well of the star.

RE 1738 exhibits shifted features in several species, at
velocities consistent with the line-of-sight ISM velocity. It is
therefore possible that these lines may be produced as a
consequence of photoionization of material passing through the
Str\"omgren sphere as the star moves through the ISM.

\subsection{G191-B2B}
We determine $V_{phot} \approx 27 \pm 2$ km/s and $V_{ism} \approx
16 \pm 1$ km/s which, considering the proximity of this star, is
unsurprisingly close to V$_{lic}$ ($\approx$ 20.6 km/s). Although
not as clearly separated as the features in RE 1738, both lines of
the \civ\ doublet are accompanied by a shifted component, the
parameters of which have been estimated by fitting multiple
gaussians to the data as illustrated in figure 1 (right). The
component at photospheric velocity is of greater equivalent width,
and the shifted elements are observed at velocities of $7.5 \pm
0.5$ km/s. However, there is no evidence for corresponding
features in any of the other lines examined.

We determine $V_{gr} \approx 15.4$ km/s from stellar models, in
agreement with the value of $19 \pm 4$ km/s estimated by Reid \&
Wegner (1988). As in the case of RE 1738, this figure is
comparable with the velocity difference between photospheric and
shifted high ionization features, suggesting that the
non-photospheric material probably resides outside the limit of
the potential well. However, the velocity is substantially
different to $V_{ism}$, and thus photoionization of the cloud
responsible for the primary ISM lines is not a viable explanation
for these features. No evidence exists for the presence of a
planetary nebula surrounding this object, but we note observations
by Green, Bowyer and Jelinsky (1990) who describe the possible
detection of a \hei\ $\lambda$584.3 emission feature in the
stellar spectrum, postulating the existence of a hidden companion
star or compact ($\sim 10$ AU) circumstellar nebula as potential
explanations for the line.

\subsection{WD 2218+706}
We estimate $V_{phot} \approx -39 \pm 1$ km/s, $V_{ism} \approx
-10$ km/s and $V_{lic} \approx 5$ km/s. Lines of the \civ\ and
\siiv\ doublets are clearly multiple, dominated by a photospheric
contribution, but with companion features of comparable equivalent
width at a velocity of $-16.5 \pm 1$ km/s. WD 2218 is surrounded
by an old planetary nebulae, and is discussed under the
alternative designation DeHt 5, by Napiwotzki and Sch\"onberner
(1995). The shifted components observed in this star are found at
a velocity redward of the photospheric lines, suggesting that this
material may be infalling; gravitational redshift 
therefore provides no viable explanation. However, Dgani and Soker
(1998) show that in regions where the ISM is reasonably dense
(such as the galactic plane), Rayleigh-Taylor instabilities can
develop in the outer regions of planetary nebulae, leading to
fragmentation of the halo, and allowing the surrounding ISM to
pass into the inner regions of the nebula where photoionization
can occur.

Although WD 2218 is out of the galactic plane ($b_{II} =
11.6$\deg), Kun (1998) describes the morphology of a nearby giant
molecular cloud complex consisting of a large number of distinct
regions previously identified in independent surveys; several are
found close to WD 2218, and two are of particular interest
(Lynds-1217 and Lynds 1219). The central portions of these clouds
have galactic coordinates within less than 0.5\deg of WD 2218, and
their distance limits (from 380 to 450 pc) encompass the distance
to WD 2218 (440 pc, from Napiwotzki \& Sch\"onberner). This raises
the possibility that the star may lie in an area where the ISM is
particularly dense, allowing instability and inflow to take place
(curve of growth analysis of the non-photospheric \civ\ features
suggests a column density of $4.17 \times 10^{13}$ cm$^{-2}$).
However, in the absence of less equivocal evidence, this
explanation must be regarded as conjecture. Alternative
explanations, such as the presence of a hidden companion, are also
deserving of investigation, and this work is currently in
progress.

\subsection{Feige 24}
Feige 24 is a white dwarf+red dwarf binary system, and several
detailed studies of this object have already been published (e.g.
Dupree \& Raymond (1982) and Vennes \& Thorstensen (1994)). Vennes
\& Thorstensen estimate a systemic velocity of $62.0 \pm 1.4$ km/s
derived from absorption line measurements. We determine $V_{phot}
\approx 30 \pm 1$ km/s, $V_{ism} \approx 3 \pm 1$ km/s and
$V_{lic} \approx 20.6$ km/s for the white dwarf. Multiple
components are observed only in the \civ\ doublet, with dominant
elements at the photospheric velocity, and secondary features at a
velocity of $5.8 \pm 1$ km/s. Our gravitational redshift estimate,
$V_{gr} \approx 17$ km/s, is somewhat higher than that derived by
Vennes \& Thorstensen ($V_{gr} \approx 9 \pm 2$ km/s), but is
still too low to explain the secondary \civ\ features. Shifted
components in the \civ\ doublet of Feige 24 have already been
discussed by Dupree \& Raymond, while Vennes \& Thorstensen
investigated the possibility that this material resided either in
the photosphere, in a circumstellar shell, or in a wind from the
red dwarf companion.

We note that the velocities of the shifted \civ\ features agree,
within error, with the ISM velocity, and hence a link between the
star and its immediate surroundings (beyond any circumstellar
shell) cannot be discounted. However, Vennes \& Thorstensen also
estimate that a \civ\ column density ranging from $8 \times
10^{11}$ to $8 \times 10^{13}$ cm$^{-2}$, (corresponding to
equivalent widths for the $\lambda$1550 line of between 4-400
m\AA), would be consistent with mass loss from the red dwarf
companion. Although insufficient data is available to derive an
unambiguous \civ\ column density in this study (figure 2 (right)),
the range of possible values we obtain includes the estimates of
Vennes \& Thorstensen. Furthermore, the equivalent width of the
$\lambda$1550 line is measured as 24 m\AA, consistent the earlier
estimates. Clearly, much work remains to be done to improve our
understanding of this system, and the recent release of a second
STIS spectrum, taken at the opposite quadrature point, is a
valuable addition to the available data.

\begin{figure}
\plottwo{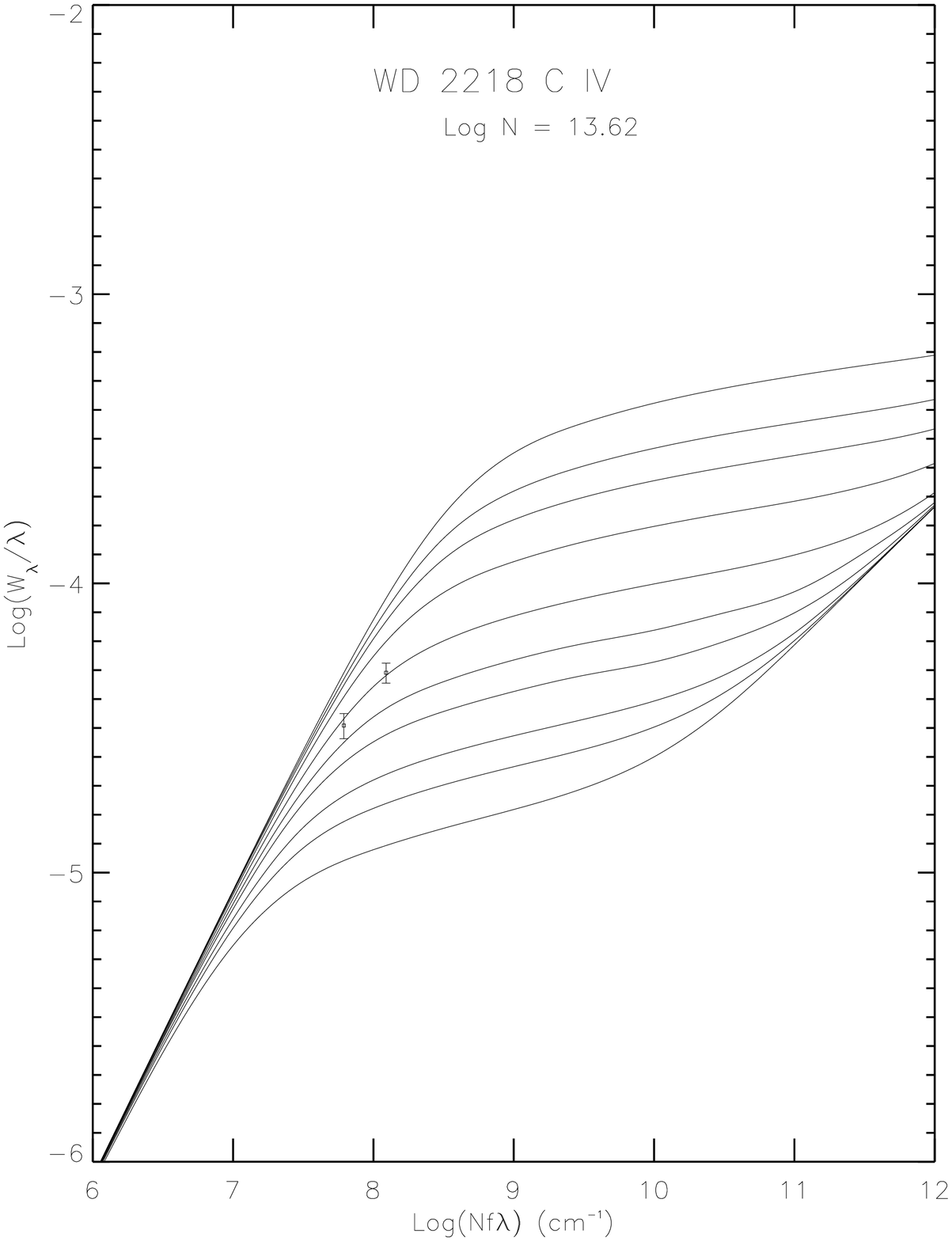}{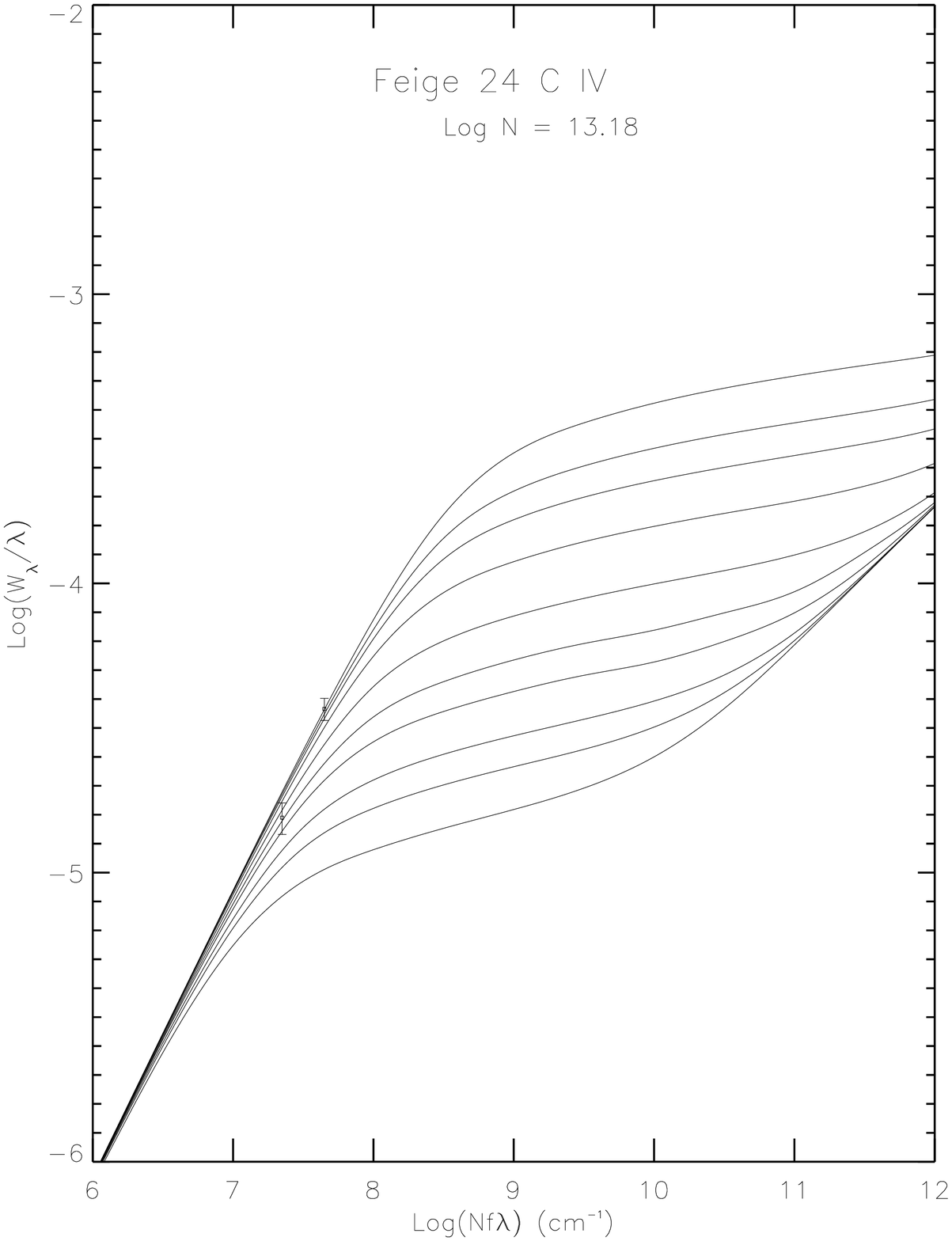} \caption{Curves of growth
for shifted features associated with the \civ\ doublet in WD 2218
(left) and Feige 24 (right).}
\end{figure}

\subsection{RE 0558+165}
We determine $V_{phot} \approx 28 \pm 2$ km/s, $V_{ism} \approx 10
\pm 3$ km/s and $V_{lic} \approx 25.2$ km/s. In common with Feige
24 and G191-B2B, this object shows evidence of non-photospheric
components only in the \civ\ doublet, at a velocity of $10 \pm 1$
km/s. Our gravitational redshift estimate suggests $V_{gr} \approx
20$ km/s, so that RE 0558 is the only star in our sample with
shifted high ionization features that could be formed by material
{\em within} the potential well, rather than the weakly shifted
outer regions. However, the highly ionized, non-photospheric
features lie at the velocity of the ISM, providing another
possible explanation for the observations.

\section{Summary}
These results represent the early stages of an ongoing study to
determine the origin of non-photospheric high ionization features
in five hot DA white dwarf stars. In RE 1738+665, the features
appear consistent with the line-of-sight velocity of the ISM, and
may be produced by photoionization of the ISM within the
Str\"omgren sphere around the star. WD 2218+706 is the only object
to show features which are redshifted with respect to the
photosphere, suggesting an infall of material. We highlight the
existence molecular clouds near this star, which may be
responsible for fragmentation of the nebula surrounding WD
2218+706, with material subsequently streaming into the
Str\"omgren sphere. The remaining three objects show shifted
features only in the \civ\ doublet. Photoionization of the ISM or
mass loss from a K-dwarf companion star are plausible explanations
in the case of Feige 24, as supported by \civ\ column density
estimates. Similar features in G191-B2B are apparently unrelated
to the ISM, and currently remain unexplained in terms of
established properties of the system. In only one case, RE
0558+165, does gravitational redshift provide more than a marginal
explanation for secondary components in the \civ\ doublet.
However, no explanation is yet available for the existence of
\civ\ features to the exclusion of other species in these three
objects.

\section{Acknowledgements}
MAB and NPB wish to thank E. Sion for his hospitality during the
2000 workshop. NPB acknowledges receipt of a PPARC studentship.

\end{document}